\documentclass[aps,prb,reprint,superscriptaddress]{revtex4-1}
\usepackage{graphicx}
\usepackage{amsmath}
\usepackage{amsfonts}
\usepackage{amssymb}
\usepackage{color}
\usepackage{bm}
\usepackage[version=3]{mhchem}
\usepackage[latin1]{inputenc}
\usepackage{siunitx}
\usepackage{epstopdf}
\pdfoutput=1

\newcommand*\bs{Bi$_2$Se$_3$}
\newcommand*\bt{Bi$_2$Te$_3$}
\newcommand*\st{Sb$_2$Te$_3$}

\begin{document}

% Use the \preprint command to place your local institutional report
% number in the upper righthand corner of the title page in preprint mode.
% Multiple \preprint commands are allowed.
% Use the 'preprintnumbers' class option to override journal defaults
% to display numbers if necessary
%\preprint{}

%Title of paper
\title{Defect Role in the Carrier Tunable Topological Insulator (Bi$_{1-x}$Sb$_x$)$_2$Te$_3$ Thin Films}

% repeat the \author .. \affiliation  etc. as needed
% \email, \thanks, \homepage, \altaffiliation all apply to the current
% author. Explanatory text should go in the []'s, actual e-mail
% address or url should go in the {}'s for \email and \homepage.
% Please use the appropriate macro foreach each type of information

% \affiliation command applies to all authors since the last
% \affiliation command. The \affiliation command should follow the
% other information
% \affiliation can be followed by \email, \homepage, \thanks as well.
%\author{}
%\email[]{Your e-mail address}
%\homepage[]{Your web page}
%\thanks{}
%\altaffiliation{}
%\affiliation{}
\author{Kane L Scipioni}
\affiliation{Department of Physics, University of Illinois Urbana-Champaign, Urbana, Illinois 61801, USA}
\author{Zhenyu Wang}
\email[]{zywang@illinois.edu}
\affiliation{Department of Physics, University of Illinois Urbana-Champaign, Urbana, Illinois 61801, USA}
\author{Yulia Maximenko}
\affiliation{Department of Physics, University of Illinois Urbana-Champaign, Urbana, Illinois 61801, USA}
\author{Ferhat Katmis}
\affiliation{Department of Physics, Massachusetts Institute of Technology, Cambridge, Massachusetts 02139, USA }
\author{Charlie Steiner}
\affiliation{Department of Physics, University of Illinois Urbana-Champaign, Urbana, Illinois 61801, USA}
\author{Vidya Madhavan}
\email{vm1@illinois.edu}
\affiliation{Department of Physics, University of Illinois Urbana-Champaign, Urbana, Illinois 61801, USA}

%Collaboration name if desired (requires use of superscriptaddress
%option in \documentclass). \noaffiliation is required (may also be
%used with the \author command).
%\collaboration can be followed by \email, \homepage, \thanks as well.
%\collaboration{}
%\noaffiliation

%\date{\today}

\begin{abstract}
Alloys of \bt{} and \st{} ((Bi$_{1-x}$Sb$_x$)$_2$Te$_3$) have played an essential role in the exploration of topological surface states, allowing us to study phenomena that would otherwise be obscured by bulk contributions to conductivity. Thin films of these alloys have been particularly important for tuning the energy of the Fermi level, a key step in observing spin-polarized surface currents and the quantum anomalous Hall effect. Previous studies reported the chemical tuning of the Fermi level to the Dirac point by controlling the Sb:Bi composition ratio, but the optimum ratio varies widely across various studies with no consensus. In this work, we use scanning tunneling microscopy and Landau level spectroscopy, in combination with X-ray photoemission spectroscopy to isolate the effects of growth factors such as temperature and composition, and to provide a microscopic picture of the role that disorder and composition play in determining the carrier density of epitaxially grown (Bi,Sb)$_2$Te$_3$  thin films. Using Landau level spectroscopy, we determine that the ideal Sb concentration to place the Fermi energy to within a few meV of the Dirac point is $x\sim 0.7$. However, we find that the post- growth annealing temperature can have a drastic impact on microscopic structure as well as carrier density. In particular, we find that when films are post-growth annealed at high temperature, better crystallinity and surface roughness are achieved; but this also produces a larger Te defect density, adding n-type carriers. This work provides key information necessary for optimizing thin film quality in this fundamentally and technologically important class of materials.
\end{abstract}

% insert suggested PACS numbers in braces on next line
%\pacs{}
% insert suggested keywords - APS authors don't need to do this
%\keywords{}

%\maketitle must follow title, authors, abstract, \pacs, and \keywords
\maketitle

% body of paper here - Use proper section commands
% References should be done using the \cite, \ref, and \label commands
\section{INTRODUCTION}
The V-VI semiconductor class of compounds contain several prototypical 3D topological insulators (TI), namely \bs{}, \bt{} and \st{}, which possess gapless spin-momentum locked surface states and an insulating bulk \cite{hasan10,qi11,zhang09}. While the existence of topological surface states has been verified \cite{chen09,xia09,hsieh09,hanaguri10,cheng10,beidenkopf11,okada11}, difficulties remain in isolating the effects of the topological states from the bulk contribution to the total conductance, which is required for further applications in electronic devices. This is attributed to inherent bulk conductivity caused by intrinsic defect doping in the binary compounds \cite{scanlon12,JiangDopingPRL}. Currently, the best way to reduce bulk carriers in \bt{} is to alloy it with \st{} \cite{zhang11,he12,he15,he13,kong11}. The rationale for this is that mixing \bt{}, which is plagued mostly by n-type Te vacancies \cite{scanlon12}, with \st{}, which mostly contains p-type anti-site impurities \cite{scanlon12,JiangDopingPRL}, using appropriate compositional ratios, will result in a net zero bulk carrier density. This method has led to the successful observations of the quantum anomalous Hall effect (QAHE) \cite{chang13} and chiral Majorana modes \cite{KWangScience17} in the thin films.

Despite these successes, a fine tuning of the chemical potential in such (Bi$_{1-x}$Sb$_x$)$_2$Te$_3$ (BST) thin films continues to remain an issue\cite{zhang11,he15,KellnerAPL}, hindering further progress. The QAHE observation, for instance, is limited to very low temperatures ($<1$K), a problem attributed to doping inhomogeneities and local chemical potential variations \cite{mogi15}. In improving film quality and performance, it is notoriously challenging to isolate and distinguish the effects of the growth parameters and composition on the properties of the thin films. For example, low temperature growth can result in lowered surface adatom mobility and create rougher films. On the other hand, growth of films at higher temperatures can result in improper nucleation\cite{harrison13}. One technique to overcome this difficulty is to grow films at lower substrate temperatures and then anneal after growth at a higher temperature \cite{harrison13,liu15}. During post-annealing, remaining adatoms as well as atoms from smaller islands are expected to become more mobile, creating larger islands and resulting in flatter films. However, post-annealing at high temperatures may also result in re-evaporation of Te and thus change the electronic properties\cite{weida2016}.

In this study, we use bulk and nanoscale characterization techniques to obtain the nanoscale morphology as well as the electronic structure of BST thin films. The composition and nanoscale structure of the thin films were determined with X-ray photoemission spectroscopy and scanning tunneling microscopy/spectroscopy (STM/S), and the powerful technique of Landau level spectroscopy was used to accurately determine the electronic properties, including the energy of the Dirac node relative to the Fermi energy. Our goal in this study is to identify the relationship between composition, growth conditions and film quality, and to particularly determine the effect of post-growth annealing on the level of disorder as well as the position of the Fermi energy with respect to the bulk bands.

\section{EXPERIMENTAL DETAILS}
BST thin films were grown using a homebuilt molecular beam epitaxy (MBE) system. The films were grown on c-plane oriented Al$_2$O$_3$ substrates, which were baked at \SI{1000}{\celsius} prior to insertion to the MBE system with a base pressure of $4\times 10^{-10}$ Torr. The growth was done by co-evaporation of Bi (99.9999\%), Sb (99.9999\%) and Te (99.9999\%) from standard single filament (Bi,Sb) and dual filament (Te) effusion cells. The substrates were held at a temperature of 180-\SI{200}{\celsius} during the growth. Typical growth rates used were 0.3-0.4 nm/min. In this letter, we compare and contrast the properties of two samples that we label sample-L and sample-H. The flux ratios were Sb:Bi = 1.36:1 and Te:(Sb,Bi)=2.1:1 for sample-L, and Sb:Bi=1.57:1 and Te:(Sb,Bi)=2.2:1 for sample-H.

Directly after growth, the films were transferred to a low temperature scanning tunneling microscope using a custom vacuum shuttle system to prevent environmental exposure. STM/S measurements were performed at 4K. In the measurements, electrochemically etched tungsten tips were used. The tunneling spectra were acquired using the lock-in technique with ac modulation about $3\;$mV at $987.5\;$Hz.

\section{X-RAY PHOTOEMISSION SPECTROSCOPY RESULTS}

\begin{figure}
\includegraphics[width=7cm]{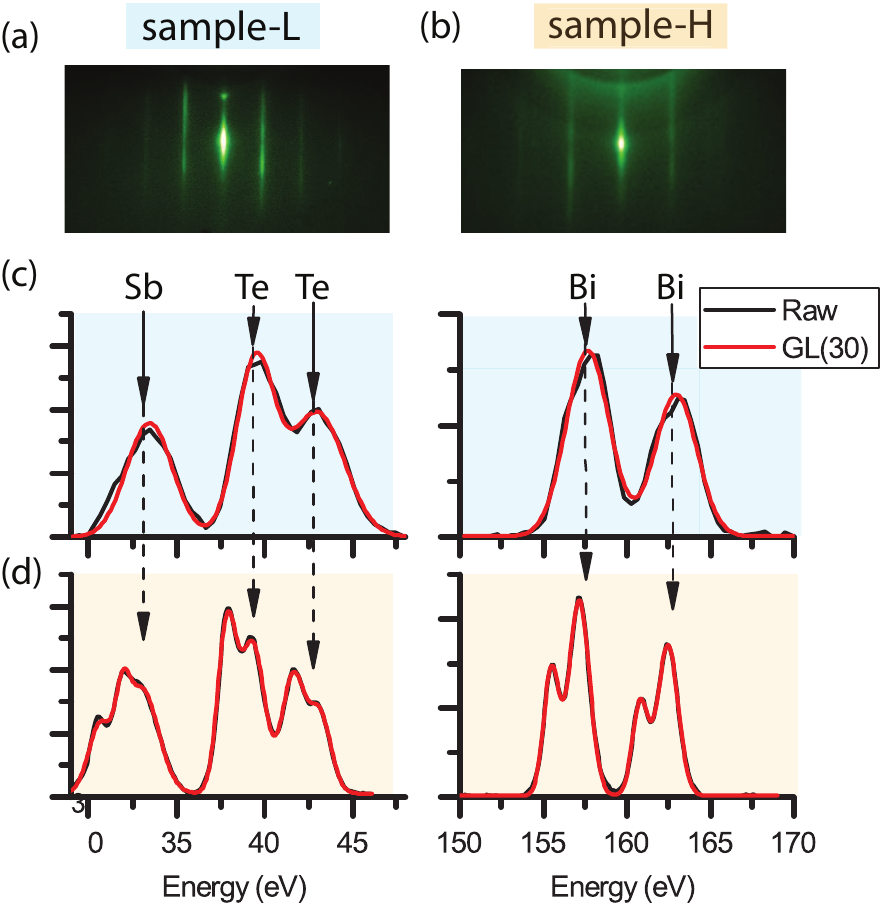}

\caption{(Color online) (a, b) RHEED patterns of sample-L (a) and sample-H (b). (c, d) The background subtracted XPS counts and the fitted mixed Gaussian-Lorentzian peaks from the sample-L (c) and the sample-H (d). The arrows indicate which elements are associated with each set of electron photo-emission peaks. The peaks are split in energy for the sample-H data because the scan step size in energy was smaller, but does not measurably affect the ratio of integrated counts.}\label{fig:rheedxps}
\end{figure}

 First, we show the RHEED patterns of two nominally similar thins films of BST, post-growth annealed at two different temperatures, \SI{220}{\celsius} (sample-L) or \SI{300}{\celsius} (sample-H) in Fig. \ref{fig:rheedxps}(a, b). Both RHEED images indicate crystalline two-dimensional film growth. The slightly sharper streaks of the RHEED patterns of sample-H potentially signify better crystallinity. This will be confirmed later when we discuss STM data on these samples.

We then determine values for Sb:Bi composition ratio, $x$, of our films. The composition was controlled by setting the effusion flux ratio of Bi and Sb during growth. However, setting a particular flux ratio is not sufficient to determine the actual composition. So, we determine the chemical composition using ex-situ X-ray photoemission spectroscopy (XPS). A Physical Electronics PHI 5400 instrument with a Mg source was used to obtain X-ray photoemission spectra as shown in Fig. \ref{fig:rheedxps} (c, d). Quantitative information was obtained by analyzing Te 4d, Bi 4f and Sb 4d emission lines. The spectra were calibrated using 1s emission line of carbon, and the peaks were fit with a Gaussian-Lorentzian product with the 30\% contribution of the Lorentzian factor (GL(30)). Quantitative information was obtained by integrating the fitted signal for all the chemical states of the chosen emission lines for each of the three elements and normalizing it with the instrument independent relative sensitivity factor (RSF), which were taken from the standard elemental library of the XPS analysis software CasaXPS. The resulting compositional fractions were $x=0.68$ for sample-L and $x=0.71$ for sample-H. For Te compositions, the Te:(Bi,Sb) ratios were 1.51 for the sample-L and 1.16 for the sample-H, indicating a Te deficiency in the latter.

\section{X-RAY DIFFRACTION RESULTS}

\begin{figure}

\includegraphics[width=6cm]{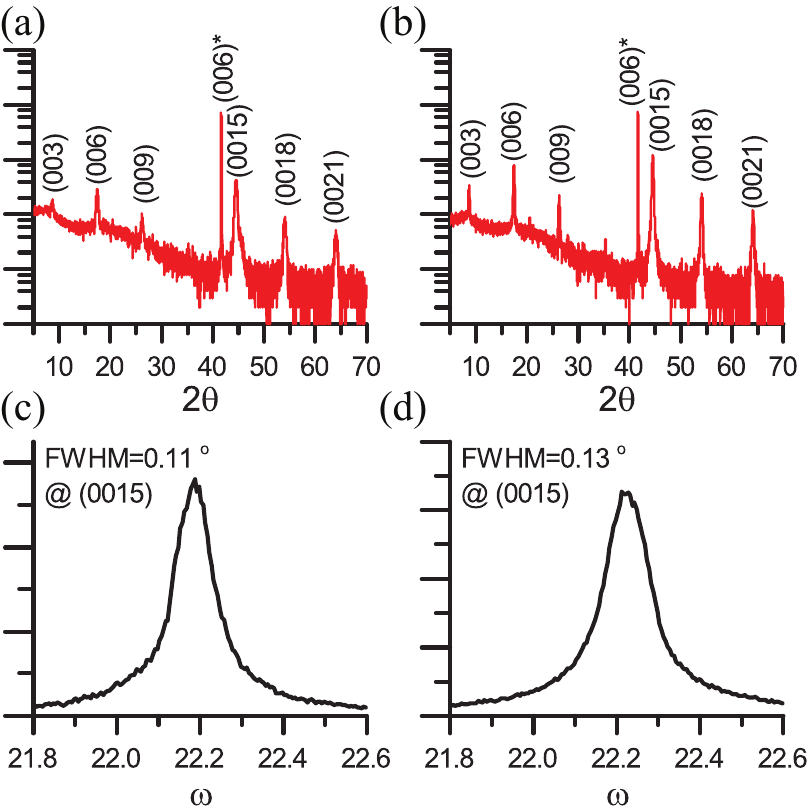}

\caption{(Color online) (a, b) $2\theta\text{-}\omega$ scans of sample-L (a) and sample-H (b). (c, d) Rocking curves of sample-L (c) and sample-H (d) about aligned with the (0015) reflection. The asterisk indicates a substrate lattice reflection.}\label{fig:xrd}
\end{figure}

To verify the absence of other crystal phases, ex-situ X-ray diffraction (XRD) measurements were performed. $2\theta$-$\omega$ coupled scans and rocking curves were obtained for both samples using a Phillips Xpert XRD system. A Cu K-alpha source was used and alignment was performed using the Al$_2$O$_3$ (006) diffraction peak. Both $2\theta$-$\omega$ scans as shown in Fig. \ref{fig:xrd}(a,b), reveal only c-plane reflections. Rocking curves, Fig. \ref{fig:xrd}(c,d), were performed about the alignment with the BST (0015) peak yielding a full-width half maximum of \SI{0.11}{\degree} and \SI{0.13}{\degree} for the sample-L and -H respectively. Taken together, the c-plane orientation, rocking curves and RHEED patterns indicate the correct crystal phase is obtained, and that the Te deficiency measured by XPS should be the result of point defects.

\section{STM/S RESULTS}

\begin{figure}

\includegraphics[width=7cm]{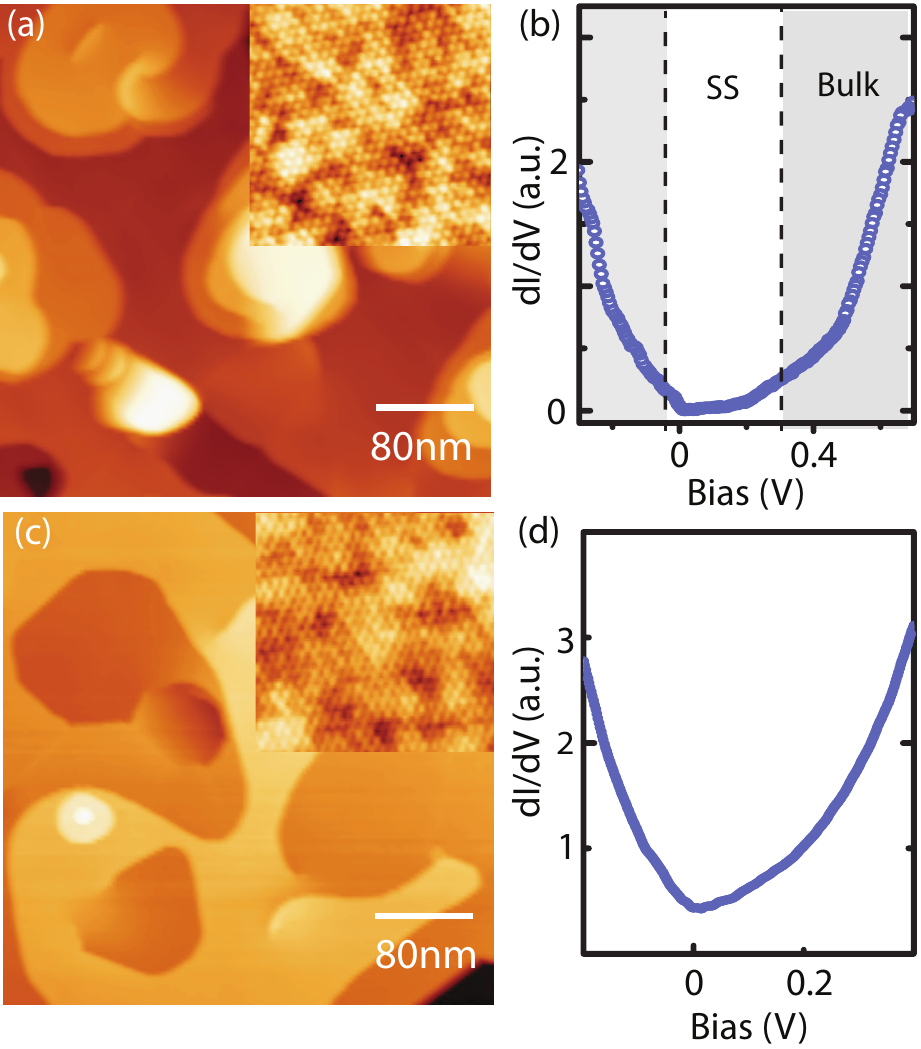}

\caption{(Color online) (a, c) 400 $\times$ 400nm and 10 $\times$ 10nm (inset) STM topographic images of sample-L (a) and sample-H (c). Images were acquired at: $V_S$ = 0.6 V, $I_t$ = 50pA for both large scans; 0.1V, 550pA for inset of (a); 0.1V, 200pA for inset of (c). (b, d) Typical dI/dV spectra of sample-L and sample-H, respectively. }\label{fig:stm}
\end{figure}

Large scale (400x400nm) topographic images of sample-L and sample-H are shown in Fig. \ref{fig:stm}(a, c). The step height between terraces is about 1 nm, corresponding to one quintuple layer. The atomic-resolution images are shown as insets and the lattice constant is about 0.435 nm. The topography shows variations at nanometer length scales, which can be attributed to the random Bi/Sb alloying.  Unlike the binary compounds\cite{JiangDopingPRL,weida2016}, it is very difficult to identify individual defects and we can not directly count the number of Te vacancies in the topographies of BST films. One possible reason is that the dominant Te vacancies lie in the middle of the quintuple layer\cite{weida2016}, and the random Bi/Sb alloy in the upper layer makes them invisible in the topography.

Though the RHEED patterns indicated reasonable crystallinity, grain boundaries and screw dislocations were observed in topographic images of both samples. The microscopic roughness can be characterized by the height of the islands within a certain area. From Fig. \ref{fig:stm}(a, c), we find that the root-mean-squared roughness of sample-L is 1.2nm and for sample-H is 0.7nm, indicating that the post-anneal temperature affects roughness at the microscale.

\begin{figure}

\includegraphics[width=9cm]{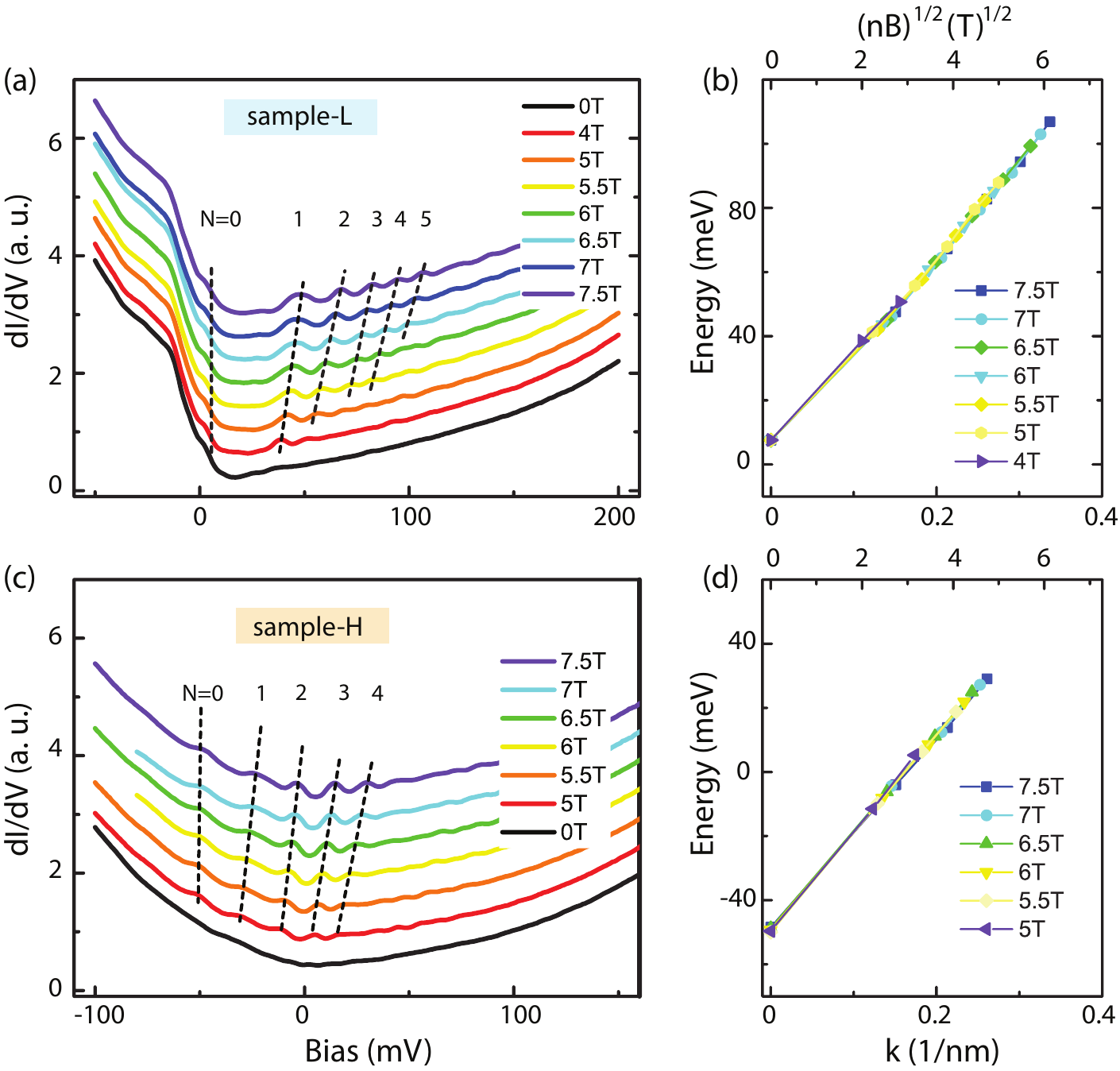}

\caption{(Color online) (a, c) Landau level spectroscopy for sample-L (a) and sample-H (c). (b, d) Dispersion of surface state for the sample-L (b) and sample-H (d). Linear fit to Eq. \ref{eq:lldisp} gives $E_D=\SI{6}{\milli\volt}$ for L and $E_D=\SI{-48}{\milli\volt}$ for H. Both sample have a Fermi velocity $v_F=4.4\times 10^5$ m/s.}\label{fig:ll}
\end{figure}

\begin{figure*}

\includegraphics[width=16cm]{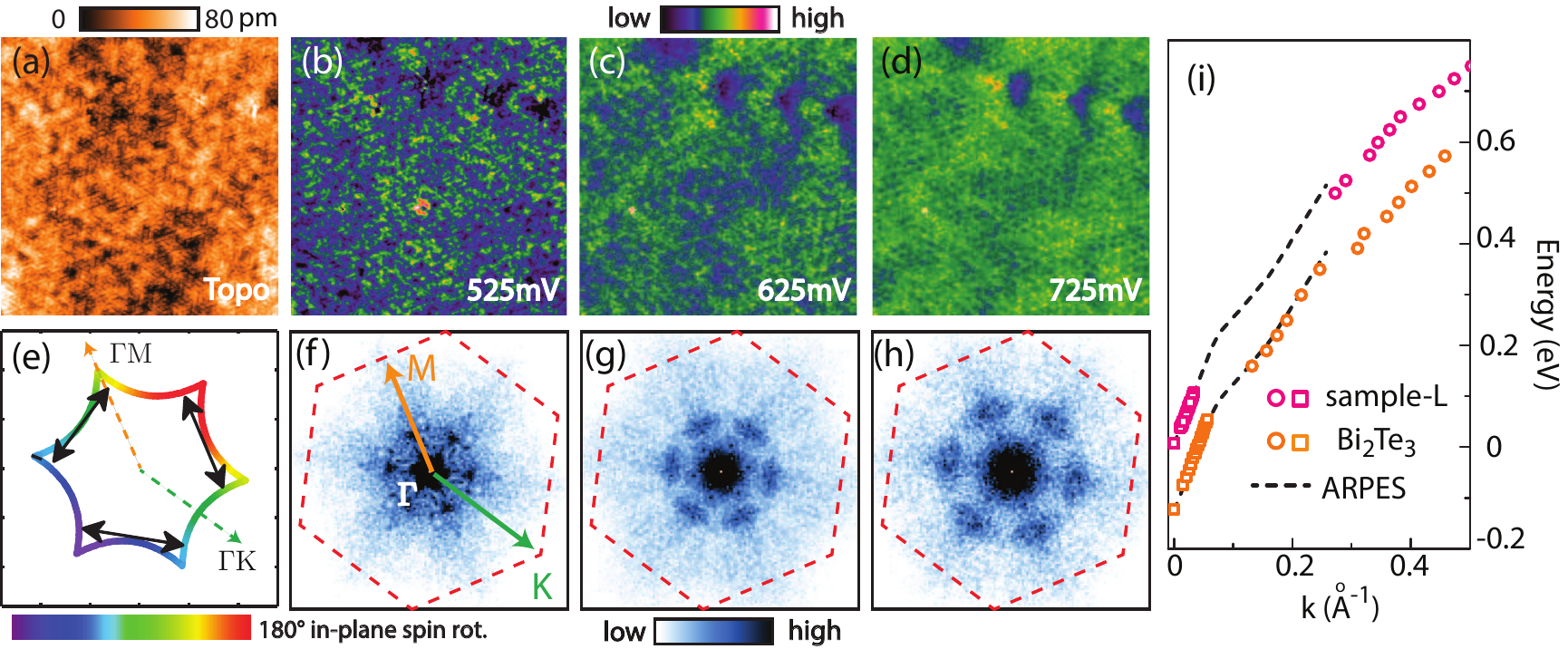}

\caption{(Color online) (a) Topography of the sample-L (40nm $\times$ 40nm). (b-d) dI/dV maps of the field of view as (a) with different energies. (e) Schematic constant energy contours in momentum space. The solid arrows denote possible scattering vectors. (f-h) FFT of the dI/dV maps shown in (b-d), respectively. The first Brillouin zone is denoted with red dash lines. (i) A summary of the dispersion of BST films and \bt{}. The orange open squares and circles are reproduced from previous LL and QPI measurements, respectively\cite{VidyaNcomm}. The dotted lines are based on previous angle-resolved photoemission spectroscopy studies of \bt{} but rigidly shifted to match the dispersion here\cite{chen09}. }\label{fig:qpi}
\end{figure*}

Next, we characterize the electronic structure and measure the energy of the Dirac point (DP) with respect to the Fermi level. The dI/dV (r, eV) spectra, which measure the local density of states (LDOS), are shown in Fig. \ref{fig:stm}(b, d). For sample-L, the LDOS is strongly suppressed from -20 meV to 260 meV, resulting in a bulk gap about 280 meV. The positions of DP, determined by the minima of LDOS, are very close to Fermi level for both samples. However, this method is fraught with problems because the tip's density of states can influence the shape of the dI/dV spectra, shifting the minimum away from the Dirac point. Additionally, the contribution of the surface states to the total density of states may be obscured by bulk bands. For example, this happens in \bt{} because the Dirac point is at a lower energy than the top of the bulk valence bands. A more accurate and reliable method to identify the position of the Dirac point is to use Landau level spectroscopy. In the presence of a magnetic field, electrons fall into quantized cyclotron orbits called Landau levels (LL), which appear as peaks in the dI/dV spectra. The energies of these states disperse with respect to magnetic field strength. For massless Dirac fermions, the dispersion is
\begin{equation}\label{eq:lldisp}
E_n=E_D+\mathrm{sgn}(n)v_F\sqrt{2eBh|n|}
\end{equation}
where $E_D$ is the Dirac point energy, $v_F$  is the Fermi velocity, $n$ is the Landau level index, and $B$ is the field strength. Assuming the usual g-factor of 2, the term resulting from the electron g-factor is negligible and has been ignored in Eq. \ref{eq:lldisp}. In fact, as we will see later, our data are consistent with this assumption. An important aspect of Eq. \ref{eq:lldisp} is that we expect to see a non-dispersing peak exactly at the Dirac point, and the energy can be identified with accuracy.

dI/dV spectra were acquired along a 15nm line cut for both samples at various perpendicular magnetic fields ranging from 0T to 7.5T. The linecut averaged spectra are shown in Fig. \ref{fig:ll} (a, c). To remove the background, the 0T spectrum was subtracted from spectra at other fields. LL peak positions were obtained by fitting the peaks to a Gaussian distribution then.  The resulting LL peak energies are plotted in Fig. \ref{fig:ll} (b, d) with respect to $\sqrt{nB}$ which can be converted into momentum. Fitting the peak positions to Eq. \ref{eq:lldisp} yields a Dirac point energy of 6 mV for sample-L and -48 mV for sample-H.  The Fermi velocity obtained from both fits is $4.4\times 10^5$ m/s. This value is consistent with previous measurements of the Fermi velocity measured for \bt{} and \st{}.

To check the unique scattering properties of the topological surface states, and obtain more information on the dispersion, we perform quasiparticle interference (QPI) measurements on sample-L. In QPI measurements, the spatial modulations in differential conductance map are recorded and then Fourier-transformed to extract scattering vectors connecting electronic states in momentum space. In Fig. \ref{fig:qpi}(c-d), we summarize the dI/dV(r, eV) maps from 500mV to 800mV, which exhibit pronounced standing wave patterns. The wavelength becomes shorter with increasing energy. These Fourier transforms after symmetrization are shown in Fig. \ref{fig:qpi}(f-h), and we observe patterns centered along the $\Gamma$M direction of the first Brillouin Zone (red dashed lines). This $\Gamma$M scattering vector originates from the hexagonally warped Dirac cones with chiral spin texture (shown as dark solid arrows in Fig. \ref{fig:qpi}(e)), and backscattering is prohibited under the protection of time-reversal symmetry\cite{ZhangQPI,FuWrap}. This measurement offers direct proof of the topological nature of the surface states. In Fig. \ref{fig:qpi}(i), we plot the dispersion along the $\Gamma$M direction together with the results obtained from LL measurements. To make a direct comparison with the dispersion of surface states in \bt{}, we show the results obtained in our previous STM study\cite{VidyaNcomm} (orange spots) and ARPES\cite{chen09} (black dashed lines) in the same plot. As can be seen, the dispersion of our BST films remains mainly unchanged, except that the Fermi level moves $\sim$130meV lower compared to the pristine \bt{}.

\section{DISCUSSION}

From our measurements, we find that sample-L is very close to the optimum composition as indicated by the position of the Dirac node, which is 6 meV from the Fermi energy. On the other hand, the Fermi level of sample-H is higher than sample-L implying it is more n-doped. Since the Sb:Bi ratio of the two samples is very similar, the difference in doping cannot be attributed to the ratio.  In fact, the slightly larger Sb content should in principle make sample-H more p-type. Given our XPS results indicate a Te deficiency in sample-H, we attribute the electron doping in sample-H to Te vacancies which are expected to act as n-type dopants in Bi/Sb rich samples. This implies that the higher annealing temperature used for sample-H caused a re-evaporation of the Te after it had been incorporated during growth.

Our results show that defects arising from the film growth conditions and the BST composition affect the Fermi level separately. Therefore, the Fermi level tuning of BST by chemical composition is not simply a matter of setting the Sb:Bi compositional ratio and depends sensitively on the post-anneal temperature. Our findings explain the large variations seen in the optimum Sb:Bi ratio of thin BST films reported by different groups \cite{zhang11,he12,he13}. In the absence of substantial Te vacancies, we find that the optimum composition for placing the Fermi energy close to the Dirac point is $x=\sim 0.7$. Moreover, our findings indicate a path to obtaining ideal samples for the QAHE. \st{} hosts a Dirac point clearly in the gap, far from the bulk bands, but is intrinsically p-type. Compensating the p-type \st{} by alloying with n-type \bt{} has been used to place the Fermi energy close to the Dirac point, but also effects the Dirac point by moving it closer to the conduction band. However, we demonstrated that post-annealing BST can introduce n-type carriers, which can be used as a parameter to obtain a finer degree of control of the electronic properties of BST films.

\section{CONCLUSION}

In summary, we have performed Landau level spectroscopy and quasiparticle interference spectroscopy, in combination with X-ray spectroscopy on BST thin films grown under different conditions. We find that Fermi energy can be placed to within a few meV of the Dirac point with Sb concentration ~0.7. However, for the same Sb/Bi ratio, the Fermi level can be tuned to an energy approximately 50meV lower simply by a higher post-annealing temperature. This explains the wide variations seen in the optimum Sb/Bi ratio of the BST films reported before, and provides key information to obtain a finer control of the electronic properties of BST films.

\begin{acknowledgments}
We would like to sincerely thank Jim Eckstein for useful discussions, and Richard Haasch for assistance with XPS data acquisition. The experiments were carried out in part in the Frederick Seitz Materials Research Laboratory Central Research Facilities, University of Illinois. V.M gratefully acknowledges support from the U. S. Department of Energy (DOE), Scanned Probe Division under award DE-SC0014335 for this work.
\end{acknowledgments}

\end{document}